\documentstyle[prd,aps,preprint,psfig]{revtex}

\def\nat{\,{Nature }}
\def\apj{\,{Astrophys. J. }}
\def\mnras{\,{Mon. Not. R. Astron. Soc. }}
\def\aap{\,{Aston. Astrophys. }}
\def\prd{\,{Phys. Rev. D }}
\def\apjs{\,{Astrophys. J. Suppl. }}
\def\araa{\,{Ann. Rev. Astron. Astrophys. }}
\def\apjl{\,{Astrophys. J. Lett. }}
\def\ptp{\,{Prog. Theor. Phys. }}

\begin{document}

\tighten
\draft
\title{
Features of Nucleosynthesis and Neutrino Emission from Collapsars
}
\author{Shigehiro Nagataki}
\address{Department of Physics, School of Science, University
of Tokyo, Tokyo 113-0033, Japan}
\author{K. Kohri}
\address{Research Center for the Early
Universe, School of Science, University of Tokyo, Tokyo 113-0033,
Japan}
\maketitle

\begin{abstract}
    In this study we present two indicators that will reflect the
    difference between collapsars and normal collapse-driven
    supernovae. They are
    products of explosive nucleosynthesis and neutrino emission.  In
    the collapsar model, it is natural to consider that the product of
    explosive nucleosynthesis depends on the zenith angle because the
    system becomes highly asymmetric in order to generate a fire ball.
    We also consider the detectability of the HNRs which is located
    nearby our Galaxy.  As a result, the number of the HNRs is
    estimated to be 5 $\times$ ($10^2$ -- $10^{-3}$), whose chemical
    composition can be spatially resolved. Using the optimistic estimate, more
    HNRs will be found and it will be possible to discuss on the
    chemical composition statistically.  As for the energy spectrums
    of neutrinos, they are not thermalized in a collapsar because the
    density of the accretion disk is much lower than that of a neutron
    star. The energy spectrums of (anti-)electron neutrinos from
    hypernovae will be mainly determined by the process of electron
    (positron) capture on free proton (neutron). It is also noted that
    high energy tail is not dumped in the case of hypernovae because
    the density of emitting region is low. Total energy of neutrino
    from hypernovae will depend on a lot of physical parameters such
    as total accreting mass and mass accretion rate, which are quite
    contrary to the situation of the normal collapse-driven
    supernovae. Therefore there will be a large variety of total
    neutrino's energies among collapsars.  In the case of SN 1998bw,
    we think that the matter around the equatorial plane might be
    ejected from the system, which resulted in the formation of
    relatively weak jets and faint GRB 980425.
\end{abstract}

\pacs{PACS number(s): 98.70.Rz, 97.60.Bw, 97.10.Cv, UTAP-364, RESCEU-13/00}

%\keywords{
%supernova: general --- supernova: individual (SN 1998bw) ---
%nucleosynthesis --- neutrino --- gamma-ray bursts
%}

%%%%%%%%%%%%%%%%%%%%%%%%%%%%%%%%
\section{Introduction} \label{intro}
%%%%%%%%%%%%%%%%%%%%%%%%%%%%%%%%
It will be a big progress that the fact that at least a part of the
gamma-ray bursts (GRBs) comes from the hypernova (HN) explosions is
being supported by the observations.  It was reported for the first
time that there seems to be a physical connection between GRB 980425
and SN 1998bw~\cite{galama98}.  They discovered an optical transient
within the BeppoSAX Wide Field Camera error box of GRB 980425. Then
they reported that the optical transient can be interpreted to be the
light curve of SN 1998bw.  As for the explosion energy of SN 1998bw,
it was estimated to be as high as (20-50)$\times 10^{51}$ erg as long
as we believe the explosion is spherically
symmetric~\cite{iwamoto98,woosley99}. This is the reason why SN 1998bw
is called as a HN. The late afterglow of GRB 970228 also suggests the
physical connection between GRB and HN~\cite{reichart99}. It was shown
that the optical light curve and spectrum of the late afterglow of GRB
970228 are well reproduced by those of SN 1998bw transformed to the
redshift of GRB 970228.  The afterglow of GRB 980326 is also believed
to be the evidence for the GRB/HN connection due to the same
reason~\cite{bloom99}.

If we believe that a part of the GRBs comes from the explosion of
massive stars, the explosion must be a jet-induced one because
spherical explosion model has a difficulty in avoiding the baryon
contamination problem~\cite{rees92}.  In fact, some observations on
GRBs are interpreted as evidence for the jet-induced explosion. For
example, the breaks in the rate of decline of several afterglows can
be explained by the beaming effect~\cite{kulkarni99}. The light curve
and spectrum of SN 1998bw also seem to suggest a jet-induced
explosion~\cite{hoflich99,nomoto_pri}.  There are also some excellent
numerical simulations on the jet-induced explosion of massive stars
whose aim is to reproduce the fire
ball~\cite{khokhlov99,macfadyen99,aloy00}, although the fire ball has
not been reproduced yet.

Here we must note the following two points. (i) it is not determined
that all of the GRBs come from HNe. (ii) the explosion energy of SN 1998bw
may be small if the explosion is the jet-induced one. Taking these points
into consideration, we can classify the relation of GRB, SN, and HN as
shown in Figure~\ref{fig1}. For example, region (a)/(d) means that
HNe which didn't/ did generate GRBs. We note that HN $\cap$ SN = $\phi$
by definition. Here we defined that SN is the explosion of a massive star
whose total explosion energy is about $10^{51}$ erg. HN is defined as the
explosion of a massive star whose total explosion energy is significantly
larger than $10^{51}$ erg. As for the region (b), other systems such as
the merging neutron stars~\cite{ruffert99} may belong to this region.

One of the most famous model to realize a GRB from a death of a
massive star is the collapsar
model~\cite{woosley93,macfadyen99,woosley99a}. The definition of the
collapsar is written in~\cite{woosley99a} as a massive star whose iron
core has collapsed to a black hole that is continuing to accrete at a
very high rate. Woosley also pointed out that there will be two types
for collapsars. One (type I collapsar) is that the central core
immediately forms a black hole with an accretion disk. The other (type
II collapsar) is that the central core forms a neutron star at first,
but the neutron star collapses to be a black hole with an accretion
disk due to the continuous fall back.  In both types, a strong jet,
which is required to produce a GRB, is generated around the polar
region due to the pair-annihilation of neutrinos that come from the
accretion disk and/or MHD processes. The remnants of a collapsar will
belong to the regions (a), (b), (c), and (d) in
Figure~\ref{fig1}. When the explosion energy of a collapsar is small,
it will be classified as SNR. When the hydrogen envelope exists, a
collapsar can not produce a GRB.

Here we note that there are no observations that support directly the scenario
of collapsars. This situation is a contrast to that of the scenario of
collapse-driven SN, which is supported by the detection of neutrinos
at Kamiokande~\cite{hirata88} and IMB~\cite{mathews88}.
Thus we present in this study two observable indicators that reflect
the mechanism of collapsars. These observations will affirm the
difference between collapsars and collapse-driven SN clearly.
These are products of explosive nucleosynthesis and neutrino emission.
We will discuss these essential features in the following sections.
We also discuss the possibility of detection of such observations
taking the event rate into consideration.

In section~\ref{nucleosynthesis}, we consider the explosive
nucleosynthesis in the collapsar model.
The luminosity and spectrum of neutrino from collapsars are shown
in section~\ref{neutrino}. Summary and discussion are presented
in section~\ref{summary}.

%%%%%%%%%%%%%%%%%%%%%%%%%%%%%%%%%%%%%%%%%%%%%%%%%%%%%%%%%%%%%%%%%%%%%%%%%
\section{Features of Remnants of Collapsars} \label{nucleosynthesis}
%%%%%%%%%%%%%%%%%%%%%%%%%%%%%%%%%%%%%%%%%%%%%%%%%%%%%%%%%%%%%%%%%%%%%%%%%
\indent

If we believe the collapsar model, the system is highly asymmetric in
order to generate a fire
ball~\cite{khokhlov99,macfadyen99,aloy00}. Thus it is natural to
consider that the product of explosive nucleosynthesis depends on the
zenith angle (see details below). So we consider its detectability in
this section.

Here we must note that the asymmetric explosion can occur in the
collapse-driven supernova as long as the effects of rotation and/or
magnetic field are taken into consideration,
e.g. see~\cite{muller80,symbalisty84,yamada94}. As a result, the
products of explosive nucleosynthesis also depends on the zenith angle
in such an asymmetric collapse-driven
supernova~\cite{nagataki97,nagataki00}. This means that we can not
distinguish well whether it is the remnant of a collapsar or of a
rotating collapse-driven supernova when we find an asymmetric SNR.  In
order to avoid such a problem, we should search for the hypernova
remnants (HNRs), whose explosion energy can not be attained by the
scenario of delayed explosion for the collapse-driven supernova.  Thus
we consider the detectability of the HNRs in this section.

To tell the truth, the products of explosive nucleosynthesis in
collapsars are not known exactly. There are many possibilities.
For example, it is pointed out that $\rm ^{56}Ni$ is synthesized by the wind
blowing off the accretion disk in a type I collapsar as long as
the disk viscosity is set to be high~\cite{macfadyen99}.
In their simulations, the outflow containing much of $\rm ^{56}Ni$ is
shown to be moving at 15 to 40 degrees off axis. However, the region
where most of $\rm ^{56}Ni$ is contained may be around the polar region
in a type I collapsar. This is the result of the explosive nucleosynthesis
behind the strong jet. This picture is like the situation which occurs
in the jet-like explosion of collapse-driven supernova~\cite{nagataki00}.
On the other hand, the chemical composition of the remnant of a type II
collapsar may be spherically symmetric, because the launch of the jet
is too late to cause the explosive nucleosynthesis~\cite{macfadyen99b}.

Such a situation is a contrast to that about the explosive nucleosynthesis
in SN. The results of numerical calculations on explosive nucleosynthesis in
collapse-driven SNe are compared with the observations
very carefully, e.g. see~\cite{hashimoto95,woosley95}.
Thus, observations of the HNRs are necessary in order to determine which
model is realistic and which model is unrealistic. Such observations may
also give a light on the occurrence frequency of the type I collapsar relative
to the type II collapsar.

Here we estimate the chance probability to find the nearby HNRs whose
chemical composition can be resolved by the latest X-ray telescopes such as
the ESA's X-ray Multi-Mirror ($XMM$) and Chandra ($AXAF$) satellites whose
spatial resolution is of order of 1 arcsec.

At first, we consider the event rate of HN in a Galaxy. If we consider
the event rate of GRB is equal to that of HN, the estimated HN rate
becomes ($10^{-6}$ -- $10^{-8}$) $\rm yr^{-1}$ per
Galaxy~\cite{cohen95,totani97,wijers98}. If we take the beaming effect
into account, the HN rate becomes larger than the observed GRB rate.
On the other hand, the HN rate can be estimated to be $\sim$ $10^{-3}$
$\rm yr^{-1}$ per Galaxy, when we assume that the slope of the initial
mass function is -1.35~\cite{salpeter55}, the maximum mass of a star
is 50$M_{\odot}$~\cite{tsujimoto95}, (10-30)$M_{\odot}$ stars
explodes as collapse-driven SNe~\cite{woosley95}, (30-50)$M_{\odot}$
stars explodes as HNe, and the collapse-driven SN event rate is
$10^{-2}$ $\rm yr^{-1}$ per Galaxy~\cite{vandenbergh91}.  This will
be an upper limit for the HN event rate because all of the massive
stars in the range (30-50)$M_{\odot}$ are assumed to explode as HNe.
So we consider that the HN rate is in the range ($10^{-3}$ --
$10^{-8}$) $\rm yr^{-1}$ per Galaxy.

In order to know the chemical composition of the ejecta, HNR must be
so young that the remnant is not composed mainly by the inter-stellar
medium (ISM) but by the HN ejecta. Here we consider the Fe distribution in the
remnant because the main products of explosive nucleosynthesis,
$\rm ^{56}Ni$, decays to Fe. We can estimate the shock radius, $R_{\rm s}$,
at which the amount of Fe from the ejecta becomes equal to that from ISM
in the remnant as follows:
\begin{eqnarray}
R_{\rm s} = 1.6 \times 10 \left( \frac{M_{\rm Fe}}{0.7 M_{\odot}} \right )
^{1/3}
\left( \frac{ 1 {\rm cm ^{-3}}  }{n} \right )^{1/3}
\left( \frac{ 1.36  }{\mu} \right )^{1/3} \;\;\; \rm  [pc],
\end{eqnarray}\label{eqn1}
where $M_{\rm Fe}$, $n$, $\mu$ are the amount of Fe from the ejecta,
mean ambient hydrogen density, and mean atomic weight of cosmic material per
H atom~\cite{allen73}. Here we assumed that the mass fraction of Fe in the
ISM is equal to that in the solar system abundances~\cite{anders89}.
On the other hand, the shock radius in the adiabatic phase can be written
as follows~\cite{sedov59}:
\begin{eqnarray}
R_{\rm s} = 7.9 \left( \frac{E_{\rm exp}}{10^{52} {\rm erg}} \right )^{1/5}
\left( \frac{ 1 {\rm cm ^{-3}}  }{n} \right )^{1/5}
\left( \frac{ t  }{ 10^3 {\rm yr} } \right )^{2/5} \;\;\; \rm  [pc],
\end{eqnarray}\label{eqn2}
where $E_{\rm exp}$ and $t$ are total explosion energy and age of the
remnant, respectively. In the case of SN 1998bw, $M_{\rm Fe}$ is
estimated to be $0.7 M_{\odot}$~\cite{iwamoto98}. If we consider that
this is the standard case with HN, $t$ has to be less than the
following value:
\begin{eqnarray}
t \leq  6.4 \times 10^3
\left( \frac{ n }{1 {\rm cm ^{-3}}} \right )^{1/2}
\left( \frac{10^{52} {\rm erg}}{E_{\rm exp}} \right )^{1/2}
\left( \frac{ R_s  }{ 16 \rm pc } \right )^{5/2} \;\;\; \rm  [yr].
\end{eqnarray}\label{eqn3} 
That is, roughly speaking, the HNR whose age is less than $10^4$ yr
must be searched for in order to know the chemical composition of the
ejecta. As for the limit of the distance from the Earth to the target,
it must be nearer than 3 Mpc in order to resolve the asymmetry of the
chemical composition of the ejecta as long as the spatial resolution
of the X-ray telescope is 1 arcsec.

Since there are 55 galaxies within 3 Mpc
from our Galaxy~\cite{tully88}, the number of the HNRs is
estimated to be 5 $\times$ ($10^2$ -- $10^{-3}$), whose chemical
composition can be spatially resolved.
($10^{-3}$ -- $10^{-8}$) $\rm yr^{-1}$ per Galaxy
and $10^4$ yr are adopted for the HN event rate and the age of the oldest
HNR, respectively. Using the optimistic estimate, the HNRs
will be found more and we will be able to discuss on the chemical
composition statistically.

Here we consider the report of Wang~\cite{wang99a} on NGC 5471B and
MF83 in M101.  They reported that NGC 5471B and MF83 may be the HNRs
since they require explosion energies comparable to the energies
frequently associated with GRBs. Since the distance of M101 from our
Galaxy is about 7.2 $\pm$ 0.4 Mpc~\cite{stetson98}, it seems difficult
to observe the distribution of the chemical composition of the
remnants at a present state. However, we can say that the HNR event
rate seems larger than the lower estimate for the GRB rate if these
are really HNRs. Although other interpretations are possible for these
highly luminous X-ray sources~\cite{chu99}, search for the hypernova
remnants nearby our Galaxy has a potential to reveal the mechanism of
the GRB.

%%%%%%%%%%%%%%%%%%%%%%%%%%%%%%%%%%%%%%%%%%%%%%%%%%%%%%%%%%%%
\section{Neutrino Emission from Collapsars} \label{neutrino}
%%%%%%%%%%%%%%%%%%%%%%%%%%%%%%%%%%%%%%%%%%%%%%%%%%%%%%%%%%%%
\indent

The second important feature of collapsars is that no neutron star but
an accretion disk around the black hole is formed. One of the most
probable heating source for the jet formation is believed to be the
$\nu \overline{\nu}$~annihilation emitted from the accretion
disk~\cite{macfadyen99}. On the other hand, neutrinos are emitted from
only the surface of a neutron star in SN explosion. In this section we
discuss the differences of the energy spectrum of the emitted
neutrinos between the collapse-driven SN and the collapsar.

In the case of SN, the energy spectrum of neutrinos is approximately
represented by the thermal distribution, because the mean free path of
neutrinos is much shorter than the radius of the neutron
star~\cite{bethe90}. Strictly compared to the perfect Fermi-Dirac
distribution with zero chemical potential, however, the high energy
phase space is less populated~\cite{janka89,myra90}. This is because
the high energy tail is dumped due to much larger opacities ($\propto$
$\epsilon_{\nu}^{2}$). In addition, it is well-known that the total
energy of neutrinos is determined by only the gravitational binding
energy of the neutron star~\cite{bethe90}.

On the other hand, the energy spectrums of neutrinos are not dumped in
collapsars, because the nucleon density of the accretion disk is much
lower than that of a neutron star~\cite{popham99}.  Namely the energy
spectrums of neutrinos emitted from collapsars are entirely
proportional to the emission rates. Then the total energy of neutrinos
could depend on a lot of physical parameters such as the total
accreting mass $M$, the mass accretion rate $\dot{M}$, and so
on. Therefore there will be a variety of total energies of the emitted
neutrino for collapsars. MacFadyen and Woosley~\cite{macfadyen99} have
shown that the accretion disk in a collapsar can be described well by
the analytic solution derived by Popham et
al.~\cite{popham99}. According to their analytic solution, the
neutrinos are mainly emitted from the region where $T$ = (1-10) MeV
and $\rho$ = ($10^9$-$10^{10}$) g $\rm cm^{-3}$.

If we assume that the density and the temperature are constant in the
neutrino emitting region~\cite{popham99}, we can estimate the energy
spectrum of the emitted neutrinos from the accretion disk.  For $n +
e^+ \rightarrow p + \bar{\nu_e}$, the spectrum of $\bar{\nu_e}$ in
unit time, unit volume, and unit energy is represented by
\begin{equation}
    \frac{d^2 n_{\bar{\nu_e}}^{eN}}{dt
    dE_{\bar{\nu_e}}}(E_{\bar{\nu_e}}) =
    \frac{G_F^2}{2\pi^3}(1+3\tilde{C_A}^2)n_nE_{\bar{\nu_e}}^2
    \sqrt{(E_{\bar{\nu_e}}^2-m_e^2)} \left(E_{\bar{\nu_e}}-Q\right)
    \frac1{e^{(E_{\bar{\nu_e}}-Q)/T}+1},
\label{eq:nusp_nuc}
\end{equation}
where $E_{\bar{\nu_e}}$ is energy of $\bar{\nu_e}$, $T$ is
temperature, $G_F$ is Fermi coupling constant, $\tilde{C}_A \simeq
1.37$ is normalized by the experimental value of neutron lifetime
$\tau_n \simeq 887.6$ s~\cite{PDG}, $n_n$ is number density of
neutron, Q $\simeq$ 1.29 MeV, and $m_e$ is electron mass. For $e^- +
e^- \rightarrow \nu_e + \bar{\nu_e}$, we obtain
\begin{equation}
    \label{eq:nusp_ann}
    \frac{d^2 n_{\bar{\nu_e}}^{e^+e^-}}{dt
    dE_{\bar{\nu_e}}}(E_{\bar{\nu_e}})=
    \frac{G_F^2}{9\pi^4}(C_V^2+C_A^2)E_{\bar{\nu_e}}^3
    \frac1{e^{E_{\bar{\nu_e}}/T}+1}T^4 \int^{\infty}_{m_e/T}
    \frac{(\epsilon^2-(m_e/T)^2)^{3/2}}{e^{\epsilon}+1}d \epsilon,
\end{equation}
where $C_V = 1/2+2\sin^2\theta_W$, $C_A$ = 1/2, and $\sin^2\theta_W
\simeq 0.231$ is Weinberg angle~\cite{PDG}, and we assume
$E_{\bar{\nu_e}} \gg T$.

In Figure~\ref{fig2}(a) we plot the obtained spectrum of $\bar{\nu_e}$
emitted from the accretion disk ($d^2n_{\bar{\nu_e}}/dt
dE_{\bar{\nu_e}} \equiv d^2n^{eN}_{\bar{\nu_e}}/dt dE_{\bar{\nu_e}} +
d^2n^{e^+e^-}_{\bar{\nu_e}}/dt dE_{\bar{\nu_e}}$) in unit time, unit
volume, and unit energy. It should be noted that the high energy tail
is not dumped at all because the nucleon density of the accretion disk
is much lower than that of a neutron star and the mean free path is
much longer. This is remarkable feature only for collapsars.
%
%---------------------------------------------
%\subsection{Total Energy}\label{flux}
%---------------------------------------------

The luminosity of $\bar{\nu_e}$ can be obtained by integrating
the spectrum as $\dot{q} \equiv  \int dE_{\bar{\nu_e}}
E_{\bar{\nu_e}} d^2n_{\bar{\nu_e}}/dt dE_{\bar{\nu_e}}$. Then we
obtain $\dot{q}^{eN} \simeq 4.6 \times
10^{33}\rho_{10}T^{6}_{11}X_{\rm nuc}\rm erg \; cm^{-3}\; s^{-1}$ and
$\dot{q}^{e^+e^-} \simeq 2.4 \times 10^{33} T^{9}_{11}\rm erg \; cm^{-3}
\; s^{-1}$,
%as below (see also
%\cite{popham99}):
%\begin{eqnarray}\label{eqn1}
%\dot{q}_{eN} \simeq 9 \times 10^{33}\rho_{10}T^{6}_{11}X_{\rm nuc} \;\;\; \rm %erg \; cm^{-3} \; s^{-1} 
%\end{eqnarray}
%\begin{eqnarray}\label{eqn2}
%\dot{q}_{\nu \bar{\nu}} \simeq 5 \times 10^{33} T^{9}_{11} \;\;\; \rm erg \; c%m^{-3} \; s^{-1} 
%\end{eqnarray}
where $\rho_{10}$ = $\rho$/$10^{10}$ g $\rm cm^{-3}$, $T_{11}$
= $T$/$10^{11}$ K, and $X_{\rm nuc}$ is the mass fraction of nucleons.
$X_{\rm nuc}$ is given by $X_{\rm nuc} = 30.97\rho_{10}^{-3/4}T_{10}^{9/8}\rm exp(-0.6096/ {\it T}_{10})$,
%\begin{eqnarray}\label{eqn3}
%X_{\rm nuc} = 30.97\rho_{10}^{-3/4}T_{10}^{9/8}\rm exp(-0.6096/ {\it T}_{10})
%\end{eqnarray}
where $X_{\rm nuc}$ $\le$ 1~\cite{qian96}.

As is clear from the above relations, the luminosity of neutrinos from
collapsars depends sensitively on the temperature. Therefore if the
configuration of the accretion disk is modified by the change of the
environment such as the mass accretion rate, mass of the progenitor,
and the mass of the black hole, then the total luminosity and energy
of neutrino from collapsars will be changed drastically. These points
are entirely different from the normal collapse-driven SN because
the total energy of neutrinos
from SN is determined only by the gravitational binding energy of the
central neutron star. The event numbers of $\bar{\nu_e}$ from
HN expected at Super-Kamiokande is represented by
\begin{equation}
    \label{eq:event}
    \frac{d R}{dE_{e^+}} =
    \frac{V_AN_p}{4\pi D^2} \sigma_{p\bar{\nu_e}}(E_{e^+})\frac{d^2
    n_{\bar{\nu_e}}}{dt dE_{\bar{\nu_e}}}(E_{e^+})\Delta t,
\end{equation}
where $E_{e^+} = E_{\bar{\nu_e}} - Q$ is the energy of the positron
which is scattered through $p + \bar{\nu_e} \rightarrow n + e^+$ in
the detector,
$\sigma_{p\bar{\nu_e}}=\frac{G_F^2}{2\pi^3}(1+3\tilde{C_A}^2)E_{e^+}\sqrt{E_{e^+}^2-m_e^2}$
is the cross section of the process, $V_A$ is the volume of the
emitting region in accretion disk, $N_p \simeq 1.5 \times 10^{35}$ is
the number of proton in Super-Kamiokande, $D$ is the distance from the
earth to the collapsar, $\Delta t \simeq M/\dot{M}$ is the duration of
the emission. In Figure~\ref{fig2}(b), we show the plot of the event
number adopting an representative parameter set~\cite{popham99}.
We can find that the neutrino emission from a collapsar
can be observed at Super-Kamiokande as long as it is located
within $\sim$3 Mpc from the earth.

As for the estimate of the detection rate at Super-Kamiokande becomes
as follows:
\begin{eqnarray}\label{eqn4}
P \sim 5\times(10^{-2} - 10^{-7}) \;\;\; \rm [yr^{-1}],
\end{eqnarray}
where the same way of estimation is done as in section~\ref{nucleosynthesis}.
That is, ($10^{-3}$ -- $10^{-8}$) $\rm yr^{-1}$ per Galaxy and 55
are adopted for the HN event rate and the number of galaxies within
3 Mpc from our Galaxy.
Using the optimistic event rate $\sim \; 5\times10^{-2}$ per year,
the detection probability of the collapsar can be as large as that of the
collapse-driven SN.

%%%%%%%%%%%%%%%%%%%%%%%%%%%%%%%%%%%%%%%%%%%%%%%%
\section{Summary and Discussion} \label{summary}
%%%%%%%%%%%%%%%%%%%%%%%%%%%%%%%%%%%%%%%%%%%%%%%%
\indent

In this study, characteristic products of nucleosynthesis and neutrino
emission have been proposed as two indicators that will reflect the
features of the collapsars.

We consider the detectability of the HNRs because we can not
distinguish well whether it is the remnant of a collapsar or of a
rotating collapse-driven supernova when we find an asymmetric SNR.  As
a result, the number of the HNRs is estimated to be 5 $\times$ ($10^2$
-- $10^{-3}$), whose chemical composition can be spatially resolved. Using the
optimistic estimate, more HNRs will be found and it will be possible
to discuss on the chemical composition statistically. Due to such
observations, we will be able to determine which model is realistic
and which model is unrealistic.  Such observations may also give a
light on the occurrence frequency of the type I collapsar relative to
the type II collapsar.  Moreover, we can say that the HNR event rate
seems larger than the lower estimate for the GRB rate if NGC 5471B and
MF83 in M101 are really HNRs. Although other interpretations are
possible for these highly luminous X-ray sources~\cite{chu99},
search for the hypernova remnants nearby our Galaxy has a potential to
reveal the mechanism of the GRB.

Strictly speaking, there will be a little difference between the SNRs
of collapsars and those of collapse-driven supernovae.  We think that
an extreme jet-induced explosion like collapsars will not happen in
the case of SN.  This is because almost all of the matter has to be
ejected in order not to leave a black hole but to leave a neutron star
at the center.  That is, matter around the equatorial plane has to be
also ejected, which will be observed as `jet-like` explosion like SN
1987A~\cite{nagataki00}. On the other hand, an extreme jet-induced
explosion is required in order to make fire balls for the model of the
jet-induced HN.  So, even if the matter around the equatorial plane is
ejected due to some reasons in the case of HN too, the degree of
jet-induced explosion will be very large and chemical composition will
depend strongly on the zenith angle in the case of the type I
collapsar.

\begin{figure}[htbp]
   \begin{center}
     \centerline{\psfig{figure=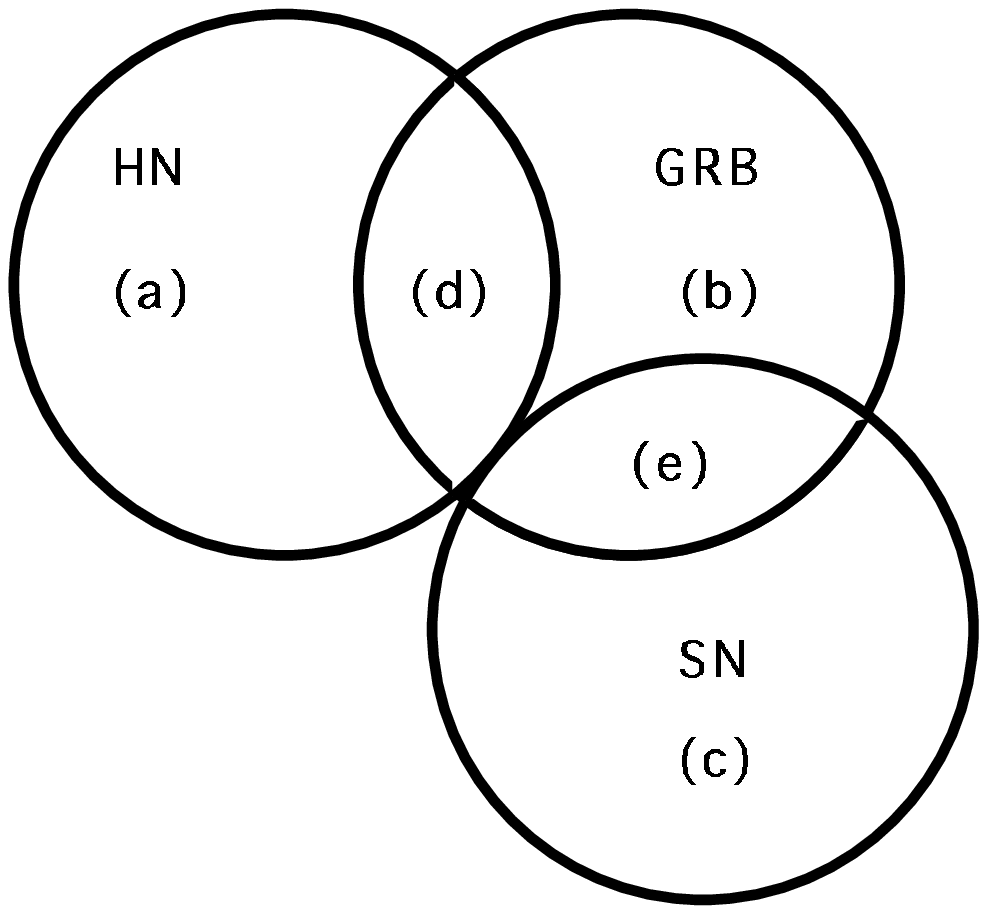,width=13cm}}
       \caption{%%
       Classification of the relation of GRB, SN, and HN. Here we
       defined that SN is the explosion of a massive star whose total
       explosion energy is about $10^{51}$ erg. HN is defined as the
       explosion of a massive star whose total explosion energy is
       significantly larger than $10^{51}$ erg. As for the region (b),
       other systems such as the merging neutron stars (Ruffert \&
       Janka 1999) may belong to this region.
       }
       \label{fig1}
   \end{center}
\end{figure}

\begin{figure}[htbp]
   \begin{center}
       \centerline{\psfig{figure=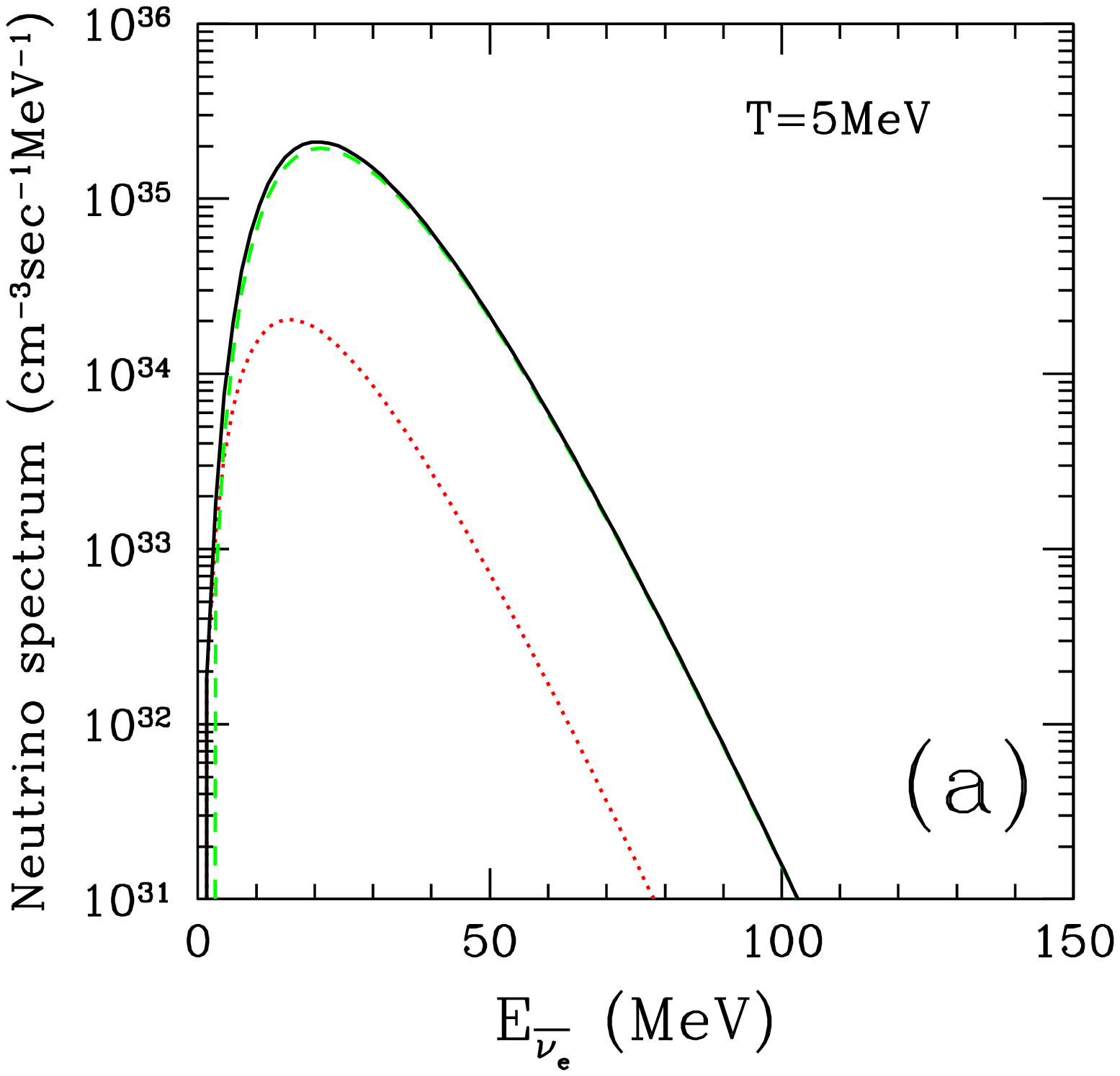,width=9cm} \psfig{figure=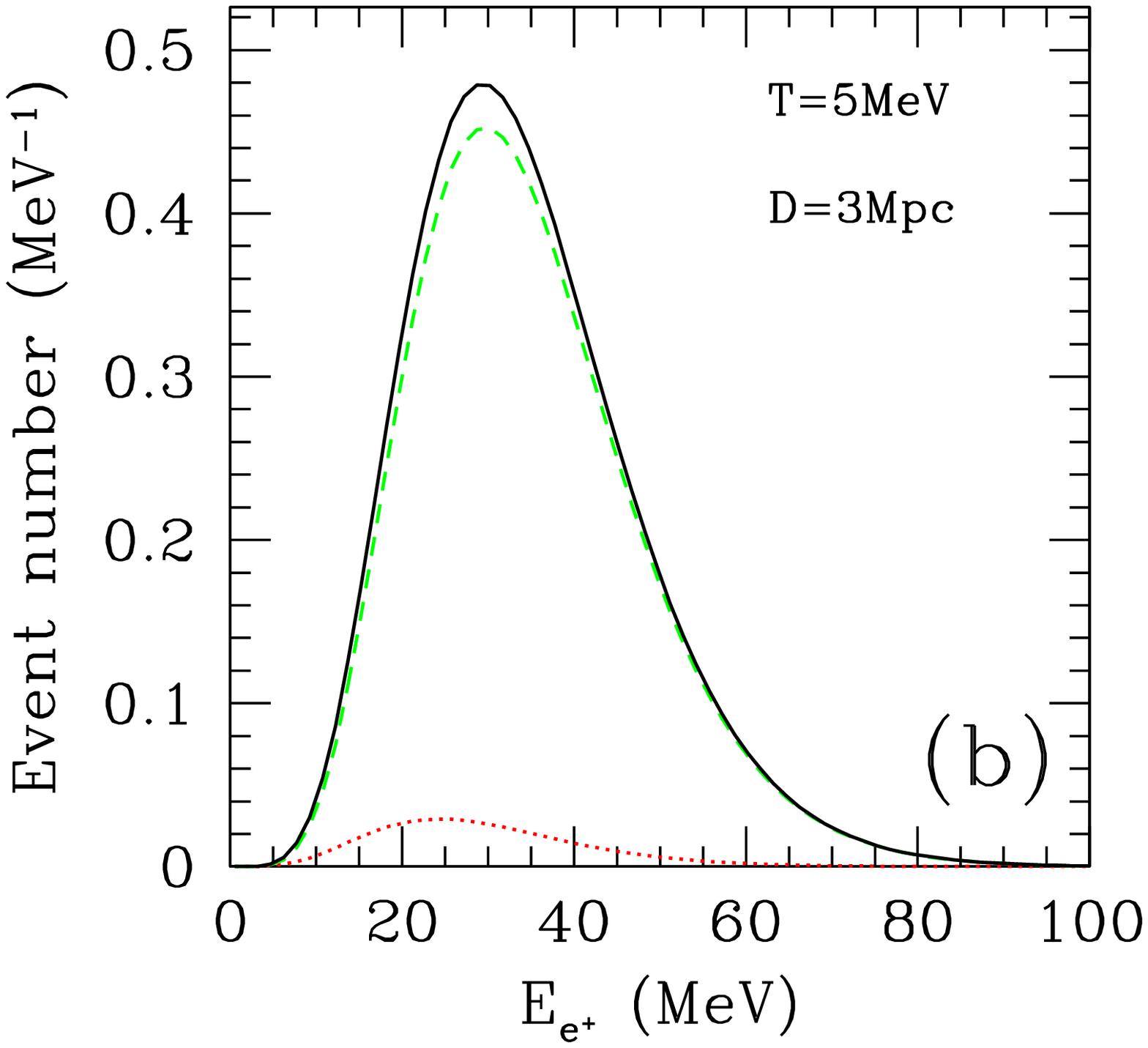,width=9cm}}
       \caption{%%
       (a) Energy spectrums of $\bar{\nu_e}$ from collapsars. (b)
       Event numbers expected at Super-Kamiokande. Solid line
       represents the total energy spectrum.  Dashed line represents
       the contribution from $n + e^+ \rightarrow p +
       \bar{\nu_e}$. Dotted line represents the contribution from $e^+
       + e^- \rightarrow \nu_e + \bar{\nu_e}$. The temperature, the
       nuclear density, and the volume of the emitting region are set
       to be 5 MeV, $10^{10}$ g $\rm cm^{-3}$, and 6.5$\times 10^{20}$
       $\rm cm^{3}$, respectively (Popham et al. 1999).  The distance,
       the total accreting mass, and the mass accretion rate are set
       to be 3 Mpc, 30$M_{\odot}$, and 0.1$M_{\odot}$ $s^{-1}$,
       respectively. Then the total event number which is obtained by
       $d E_{e^+}$ integration is $\sim$ 15.
       }
    \label{fig2}
   \end{center}
\end{figure}

It is also noted that the mass accretion rate becomes low if the
matter around the equatorial plane is ejected from collapsars.  This
will result in the decline of the total energy of neutrinos emitted
from the accretion disk. As a result, total explosion energy may
become small in that case.  It is reported that the explosion energy
of GRB 980425, which is said to be associated with SN 1998bw, is quite
lower than that of the usual GRBs~\cite{galama98}. In the case of SN
1998bw, we think that the matter around the equatorial plane might be
ejected from the system, which resulted in the formation of relatively
weak jets and faint GRB 980425. This means that SN 1998bw and GRB
980425 may be classified in the region (e) in Figure~\ref{fig1}. Of
course, this picture requires that the system of SN 1998bw and GRB
980425 is highly asymmetric, because the total explosion energy of SN
1998bw is estimated to be (20-50)$\times 10^{51}$ when spherical
explosion is assumed~\cite{iwamoto98,woosley99}.

As for the (anti-)electron neutrino emission from the collapsars,
its energy spectrum is mainly determined by the emission rate due
to electron (positron) capture on proton (neutron). As the temperature
becomes higher, contribution of the process of electron-positron
pair annihilation can not be negligible. It is also noted that
high energy tail is not dumped in the case of the collapsar
because the density of emitting region is low. These features
on energy spectrum are quite different from that of SN.

Total energy of neutrino depends on many physical quantum such as
total accreting mass and mass accretion rate. It is noted the
emission rate due to the electron capture on proton is proportional
to $T^{6}$. So a little change in temperature results in great
change in the neutrino flux. That is why there will be a variety of
total luminosity of neutrino among collapsars, which is in striking
contrast to the case of SN. As for the event rate,
the detection probability of the collapsar can be as large as that of the
collapse-driven SN if we use the optimistic event rate
$\sim \; 5\times10^{-2}$ per year at Super-Kamiokande.

Finally, we stress again that these features on nucleosynthesis and
neutrinos will reveal the mechanism of GRB quite well. We hope the
increase of further observations in the near future.

\acknowledgements
This research has been supported in part by a Grant-in-Aid for the
Center-of-Excellence (COE) Research (07CE2002) and for the Scientific
Research Fund (199908802, 199804502) of the Ministry of Education,
Science, Sports and Culture in Japan and by Japan Society for the
Promotion of Science Postdoctoral Fellowships for Research Abroad.

%%%%%%%%%%%%%%%%%%%%%%%%%%%%%%%%%%%%%%%%%%%%%%%%%%%%%%%%%%%%%%%%%%%%%%

%%
\newpage

\newpage

\end{document}